# Surface salinity fields in the Arctic Ocean and statistical approaches to predicting anomalies and patterns

Ekaterina Chernyavskaya, Department of Oceanography, Arctic and Antarctic Research Institute, St. Petersburg, Russia.

Ivan Sudakov, Department of Mathematics, University of Utah, Salt Lake City, Utah, USA.

Kenneth M. Golden, Department of Mathematics, University of Utah, Salt Lake City, Utah, USA.

Leonid Timokhov, Department of Oceanography, Arctic and Antarctic Research Institute, St. Petersburg, Russia.

*Corresponding author:* I. Sudakov, Department of Mathematics, University of Utah, 155 S 1400 E, Room 233, Salt Lake City, UT 84112-0090 USA. (sudakov@math.utah.edu)

**Key Points:**

- Statistical approaches to analyzing the Arctic Ocean salinity were developed.
- Six kinds of typical patterns in the surface salinity fields were identified.
- Abrupt changes in the Arctic Ocean surface layer salinity were found.

**Abstract**

Significant salinity anomalies have been observed in the Arctic Ocean surface layer during the last decade. Using gridded data of winter salinity in the upper 50 m layer of the Arctic Ocean for the period 1950-1993 and 2007-2012, we investigated the inter-annual variability of the salinity fields, attempted to identify patterns and anomalies, and developed a statistical model for the prediction of surface layer salinity. The statistical model is based on linear regression equations linking the principal components with environmental factors, such as atmospheric circulation, river runoff, ice processes, and water exchange with neighboring oceans. Using this model, we obtained prognostic fields of the surface layer salinity for the winter period 2013-2014. The prognostic fields demonstrated the same tendencies of surface layer freshening that were observed previously. A phase portrait analysis involving the first two principal components exhibits a dramatic shift in behavior of the 2007-2012 data in comparison to earlier observations.

## 1. Introduction

The Arctic Ocean is very sensitive to changing environmental conditions. Its surface layer is a key component of the Arctic climate system, which constitutes the dynamic and thermodynamic link between the atmosphere and the underlying waters [*Carmack*, 2000]. The stability and development of the ice cover are associated with mixed layer thickness, upper layer salinity and upper halocline, which influence the geographic distribution of sea ice and its variability. In this context, the Arctic Ocean surface layer is a critical indicator of climate change in the Arctic [*Zaharov*, 1996].

The thermohaline structure of the Arctic Ocean surface layer has undergone significant changes in recent years (Figure 1). Of particular interest is the great freshening of the Canada Basin surface layer that had not been observed since 1950 [*Timokhov et al.*, 2011] until the early 1990s. However, in *Jackson et al*. [2012] it was emphasized that processes related to warming and freshening the surface layer in this region, had transformed the water mass structure of the upper 100 m.

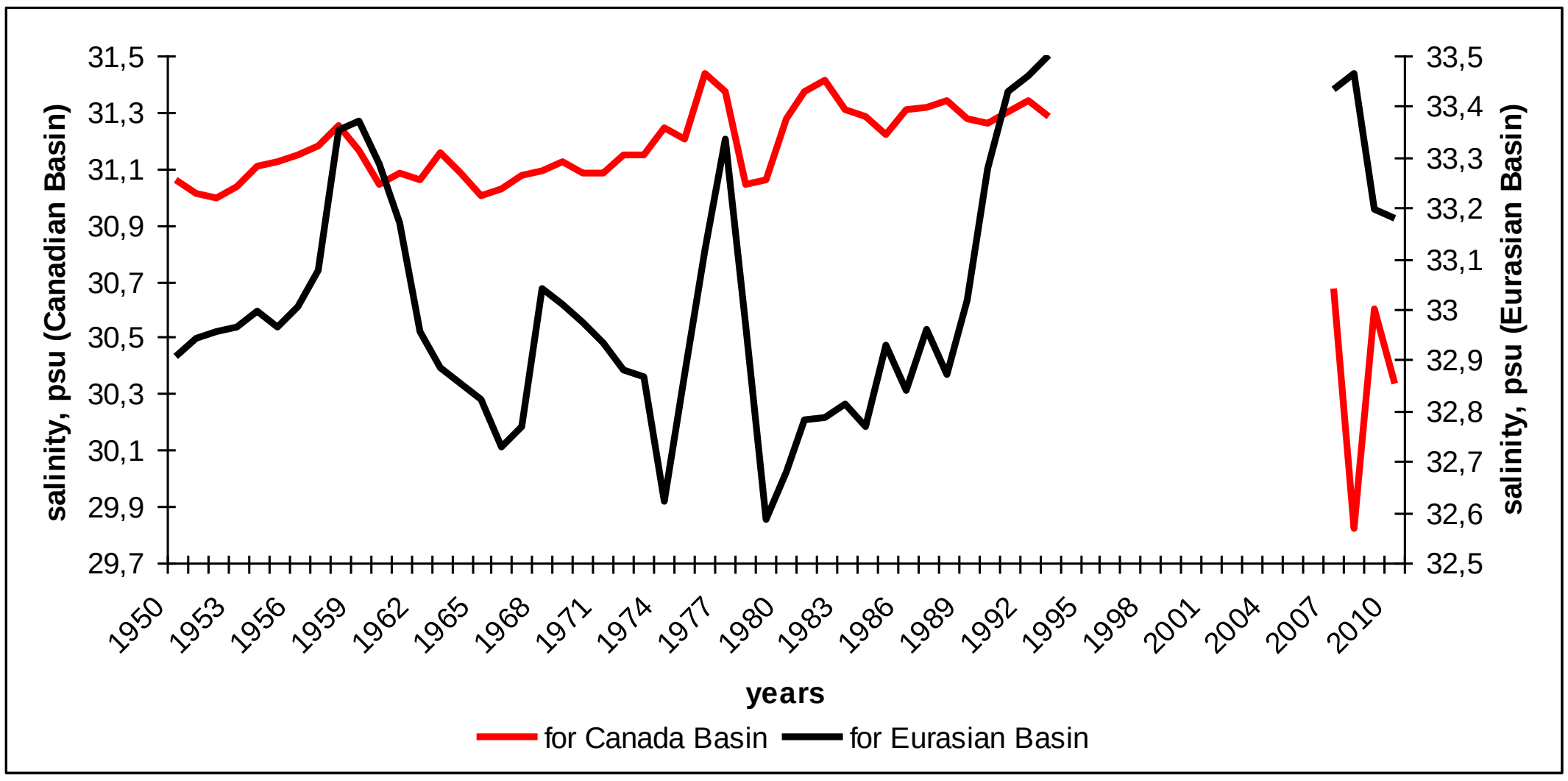

*Figure 1. Temporal changes in salinity averaged over the depth range 5-50 m. Data are shown for the Eurasian and Canada Basins, and display some degree of anomalous behavior.*

In addition, there have been observations of significant salinification of the upper Eurasian Basin that began around 1989. One hypothesis for this is that the increase of Arctic atmospheric cyclone activity in the 1990s led to a large change in the salinity in the Eurasian Basin. This can be explained through two mechanisms of salinization: 1) changes in river inflow, and 2) increased brine formation due to changes in Arctic sea ice formation. The high salinization in this region altered the formation of cold halocline waters, weakened vertical stratification, and released heat upward from below the cold halocline layer [*Johnson and*

*Polyakov*, 2001]. The other reason for salinification is the influence of the Atlantic waters (AW), which by 2007 became warmer by about 0.24˚C than they were in the 1990s. Observations show that increases in Arctic Ocean salinity have accompanied this warming. This led to significant shoaling of the upper AW boundary (up to 75-90 m in comparison with climatic values) and the weakening of the upper-ocean stratification in the Eurasian Basin as well [*Polyakov et al.*, 2010]. However, current observations also show that the upper ocean of the Eurasian Basin was appreciably fresher in 2010 than it was in 2007 and 2008 [*Timmermans et al.,* 2011].

In *Zhange et al.* [2003] it is emphasized that the fresh water balance and salinification of the Arctic Ocean are important factors for processes in the mixed layer. In turn, it is well known that the crucial factors of the surface water mass transformation are advection of the salinized ocean waters, the influence of this process on the halocline, and, on the other hand, changes in the density field, conduction to the surface, and water and sea ice circulation.

It is known that for the Arctic Ocean, water density depends more on water salinity than on water temperature, and hence the thermohaline circulation is mainly determined by salinity distribution. This conclusion comes easily from an analysis of a linear equation for the state of sea water:

$$\rho = \rho_0 - \varepsilon T(T - T_0) + \varepsilon S(S - S_0), \qquad (1)$$

where $\rho_0$, $T_0$, $S_0$ are some initial values of water density, temperature, and salinity;
and $\varepsilon T = 7\cdot10^{-5}$ g/(cm$^3$·K), $\varepsilon S = 8\cdot10^{-4}$ g/(cm$^3$·‰).

Vertical variations of temperature and salinity in the upper layer can reach 0.5˚C and 1‰ respectively. If we put these numbers into equation (1) we can obtain the contributions of variations in temperature and salinity to changes in water density, which are about 4% and 96%, respectively. Thus, salinity was chosen as the main characteristic of thermohaline structure variations of the Arctic Ocean surface layer.

The transfer from the Arctic Ocean to the North Atlantic of briny water from sea ice formation and sea ice itself are significant components of global ocean circulation. Thus, the investigation of the variability of the surface layer can make a significant contribution to understanding ocean-climate feedback. Particularly, abrupt changes in surface layer salinity may lead to a so-called *tipping point* in global ocean circulation [*Lenton et al.*, 2009]. A “tipping point” [*Lenton*, 2011] may occur if a small change in forcing triggers a strongly nonlinear change of the internal properties of the system, leading to a change in its future state. We may interpret “a small change in forcing triggers” as anomalies in inter-annual salinity variability. Robust models are required for implementation of this hypothesis. At the present time we have a number of different physical models of surface layer salinity. For example, sea ice salinity models can describe significant changes in physical macroscopic properties, as well as microscopic properties such as the distribution of brine channels [*Vancoppenolle et al.*, 2009]. Using regional climate models (for specific seas) for the understanding of scale variation is perhaps not an appropriate approach. A robust statistical model may help to describe features of anomalies in salinification of the Arctic Ocean, which are key players in the formation of surface salinity patterns. In this case, investigation of the structure of patterns and quality of anomalies leads to a better understanding of possible tipping points in global ocean circulation.

Variations of Arctic Ocean surface salinity have complex spatial and temporal structures, which are affected by many external factors. Our aim is to distinguish the most significant factors that lead to recent changes in upper layer salinity patterns.

We propose to develop our model in view of the ideas of *Timokhov et al.* [2012]. This statistical model of variability of the Arctic Ocean winter salinity in the 5–50 m layer is used as a method of reconstruction of the winter salinity fields, which have been suggested in *Pokrovsky and Timokhov* [2002]. The model is based on equations of multiple correlations for the time series - principal components (PC) associated with the first five leading modes of the Empirical Orthogonal Function (EOF). We apply this type of spectral analysis to the salinity fields. The contribution of atmospheric factors, hydrological processes, and pre-history of the spatial distribution of salinity can be interpreted through determining the structure of the multiple correlation equations.

Based on gridded data of winter salinity of the upper 50 m layer for the periods of 1950-1993 and 2007-2012, we have investigated features of inter-annual variability of the salinity fields, attempted to identify the causes of its anomalies, and made a statistical model for the prediction of surface layer salinity fields.

Cluster analysis of the surface salinity allowed identification of six types of patterns in the salinity fields, which differ from each other by the position of the fresh water core, the position of the Transpolar Drift frontal zone, and the value of the horizontal salinity gradient. It has been shown that the structure of the salinity fields of 1990-1993 and 2007-2012 differed greatly from previous years. The uniqueness of the haline structure (during 2007-2012) was also confirmed by the results of decomposing the surface salinity fields into Empirical Orthogonal Functions.

Analysis of the equations for the first five PCs showed that surface salinity fields were influenced most by atmospheric processes. Moreover, the pattern structure of the salinity fields due to their conservative nature can store and accumulate the after-effects of atmospheric processes occurring 2-3 years ago – according to the results of our correlation analysis of the links between principal components and various external factors.

Using the PCs calculated by the model, we obtained forecast fields of the Arctic Ocean surface layer salinity for the winter period 2013-2014. Prognostic fields demonstrate the same tendencies of the surface layer freshening that were previously observed.

## 2. Data Sets and Method

### 2.1. Data Set

This study is based on the collection of more than 6,419 instantaneous temperature and salinity profiles with data available at the standard levels (5, 10, 25, 50, 75, 100, 150, 200, 250, 300, 400, 500, 750, 1000 and so on every 500 meters) collected between 1950-1993 and obtained from the Russian Arctic and Antarctic Research Institute (AARI) database. This is complemented by data made available between 2007-2012 from the expeditions of the International Polar Year (IPY) and after which consisted of Conductivity Temperature Depth (CTD) and eXpendable Conductivity Temperature Depth (XCTD) data originating from the Ice-Tethered Profiler (ITP)-buoys. The average vertical resolution of these profiles was 1 m. The first database was introduced by *Lebedev et al.* [2008]. In areas where observations were missing, temperature and salinity data were reconstructed in a regular grid for the period of 1950 to 1989. Also, some data was found in the joint U.S. Russian Atlas of the Arctic Ocean for winter [*Timokhov and Tanis*, 1997]. Thus, the working database is represented by grids with spatial resolutions of grid points 200 per 200 km, covering a deep part of the Arctic Ocean (with depths of more than 200 m). According to *Treshnikov* [1959] and *Rudels et al.* [1996, 2004], the average thickness of the Arctic Ocean mixed layer for the winter season is 50 m. Thermohaline characteristics of the surface layer fully reflect the effects of atmospheric and ice processes, as the water most directly exposed to the atmosphere and ice lie within the mixed layer [*Sprintall and Cronin*, 2001]. Also, a description of the data sources for other physical parameters can be found in Table 1.

### 2.2. Gridding algorithm

Usually a long-term continuous series of observations are required for statistical analysis. However, the Arctic Ocean temperature and salinity data may be unrepresentative because the network of the stations cannot cover all observation areas. Thereby, we will provide analysis of the inter-annual variability of the surface layer salinity. With this aim, we have unified existing data sets using reconstruction and gridding. The technique of computation of gridded fields for the period from 1950-1993 was described by *Lebedev et al.* [2008].

These techniques are based on the method of ocean field reconstruction, proposed by *Pokrovsky and Timokhov* [2002]. This method, which was used to obtain gridded salinity fields, is described as

$$z_i = z_i^{(r)} + e,\ \langle z_i z_j \rangle = \sigma_{x_i x_j},\ \langle z_i e_i \rangle = 0, \tag{2}$$

$$\langle e_i \rangle = 0,\ \langle e_i e_j \rangle = \delta_{ij}\sigma_e^2.$$

We assume that $z(t, x)$ is the measured value of an oceanographic parameter (e.g. temperature or salinity) with a standard deviation σ and is a random function of time $t$ and spatial coordinates $x \in \Re$. We can reproduce the observed value of $z(t, x)$ as the sum of a true value $z^{(r)}(t, x)$ of the oceanographic parameter and an observational error $e$ $(t, x)$. We also propose that $z_i^{(r)}$ has spatial correlations to the parameters; that a systematic error is not identified; a standard deviation of error ($\sigma_e^2$) does exist and $\delta_{ij}$ is the Kronecker delta.

Biorthogonal decomposition of the oceanographic record can help to identify the connections between spatial and temporal distributions within the data:

$$z(t_j, x_i) = \sum_k c_k^j f_k(x_i) + e(t_j, x_i), \tag{3}$$

where $f_k(x_i)$ – is the spatial empirical orthogonal function (EOF); $c_k^j$– the calculated coefficient, so-called ***k*-th** principal component.

As the next step we approximate the EOF through linear combinations of convenient analytical functions $P_l\,(x_i)$. Thus, the modified biorthogonal decomposition can be written:

$$z(t_j, x_i) = \sum_k d_l^j P_l(x_i) + e(t_j, x_i), \tag{4}$$

and here as $d_l^j = \sum b_{kl} c_k^j$ .

The main goal of this spectral analysis method is to estimate coefficients of spectral decomposition $C = \left| c_k^j \right|$ or $B = \{ b_{kl} \}$. Actually, this approach is a combination between singular value decomposition and statistical regularization. These coefficients (modes) can be marked through the real physical processes which influence salinity (see Table 1).

Preparation of the average salinity field data for 2007-2012 has consisted of several stages. First, we have checked data for random errors. Further, we have used linear interpolation, as well as assimilated the real plane with the field data through the virtual plane of data. Next, we constructed an interpolation via a grid of nodes (separately for each plane). The gaps in the data for uncovered sites were filled from the Joint U.S.-Russian Atlas of the Arctic Ocean [*Timokhov and Tanis*, 1997].

Table 1. Predictors used for the approximation of PCs.

| **Physical processes and its notation** | **Physical value** | **Description** | **Data sources (references and the Internet sources)** |
|---|---|---|---|
| Arctic oscillation index (AO) | Sea-level pressure anomaly north of 20N latitude | When the AO index is positive, surface pressure is low in the polar region.<br>When the AO index is negative, there tends to be high pressure in the polar region.<br>Here and further, the lower case indicates the months of an average period. | *Zhou et al.* [2001].<br>NOAA Center for Weather and Climate Prediction (NCWCP)<br>http://www.cpc.ncep.noaa.gov/ |
| North Atlantic oscillation index (NAO) | Sea-level pressure anomaly between the Icelandic low and the Azores high | When the NAO index is positive, pressures in the Azores high are especially high and pressures in the Icelandic low are lower than normal. Both pressure systems are located to the north.<br>When the NAO index is negative, the Azores high and the Icelandic low are much weaker. Pressure differences are therefore smaller and both systems are located to the south. | *van den Dool et al.* [2000].<br>NOAA Center for Weather and Climate Prediction (NCWCP)<br>http://www.cpc.ncep.noaa.gov/ |
| Pacific/North American index (PNA) | Sea-level pressure anomaly in the Northern Hemisphere extra tropics | When the PNA index is positive, above -average heights over Hawaii and over the intermountain region of North America, and below-average heights located south of the Aleutian Islands and over the southeastern United States.<br>When the PNA index is negative, strong and extensive Hawaii high and a weak and very local Aleutian low are observed. | *Chen et al.* [2003].<br>NOAA Center for Weather and Climate Prediction (NCWCP)<br>http://www.cpc.ncep.noaa.gov/ |
| Arctic Dipole Anomaly index (DA) | Sea-level pressure anomaly north of 20N latitude | When the DA index is positive, sea-level pressure has positive anomaly over the Canadian Archipelago and negative anomaly over the Barents Sea.<br>When the DA index is negative, SLP anomalies show an opposite scenario, with the center of negative SLP anomalies over the Nordic seas. (Wu et al., 2006; Wang et al., 2009; Overland & Wang, 2010). | *Kistler et al.* [2001].<br>Joint Institute for the Study of the Atmosphere and Ocean (JISAO)<br>http://www.jisao.washington.edu/analyses0302/ |
| Atlantic Multidecadal oscillation index (AMO) | Variations of sea surface temperature in the North Atlantic Ocean | Index has cool and warm phases that may last for 20-40 years at a time and a difference of about 0.5˚C. It reflects changes of sea surface temperature in the Atlantic Ocean between the equator and Greenland.<br>Was used as a substitute for processes of water exchange with the Atlantic Ocean. | *Enfield et al.* [2001].<br>ESRL Physical Sciences Division (PSD)<br>http://www.esrl.noaa.gov/psd/data/timeseries/AMO/ |
| The Pacific Decadal Oscillation index (PDO) | North Pacific sea surface temperature variability | When the PDO index is positive, the west Pacific becomes cool and part of the eastern ocean warms. When the DA index is negative, the opposite pattern occurs. It shifts phases on at least the inter-decadal time scale, usually about 20 to 30 years. | *Trenberth and Hurrell* [1994]. Joint Institute for the Study of the Atmosphere and Ocean (JISAO) http://jisao.washington.edu/pdo/ |
| Air temperature anomaly ($T_{air}$) | Degree | Monthly average anomalies of air temperature over the Arctic. The lower case indicates months of an average period. | *Smith* [2011]. NOAA's National Climatic Data Center (NCDC) http://www.ncdc.noaa.gov/cag/time-series/ |
| River runoff (RIV) | Water flows | Average annual runoff of the main Siberian rivers. It was used as total runoff in the Kara Sea (K), Laptev Sea (L), East-SiberianSea (E) and Chukchi Sea (C). The lower case indicates the first letters of the sea name. | *Timokhov and Tanis* [1997]. Joint US-Russian Atlas of the Arctic Ocean. http://rims.unh.edu/data/station/list.cgi?col=4 |
| Ice extent (IceExt) | Area | Area of ice extent in the Arctic Ocean in September. | *Cavalieri et al.* [1992]. ESRL Physical Sciences Division (PSD) http://www.esrl.noaa.gov/psd/data/gridded/tables/arctic.html |
| Area of open water in Arctic seas (OW) | Area | Total ice-free area in the Kara Sea (K), Laptev Sea (L), East-Siberian Sea (E) and Chukchi Sea (C) in September. The lower case indicates the first letter of the sea name. | *Lebedev et al.* [1992]. Russian Arctic and Antarctic Research Institute (AARI) http://www.aari.ru/projects/ECIMO/index.php?im=100 |
| Bering Strait inflow (BS) | Water flows | Average annual water exchange through the Bering Strait. | *Roach et al.* [1995]; *Woodgate et al.* [2005]. |

### 2.3. Statistical approach

Here we describe the approaches to data analysis which were used for physical interpretation of our statistical model.

Researchers *Polyakov et al.* [2010]; *Rabe et al.* [2011]; *Morison et al.* [2012] have emphasized that the thermohaline structure of the surface layer has undergone significant change over the last decade. However, we still don't understand the physical processes which led to these changes or what might be the future trends.

On the other hand, we can assume that the analysis of the variability of the surface layer (including salinity fields) of the Arctic Ocean may be based on the decomposition using empirical orthogonal functions. This approach is useful in our case because decomposition by EOF gives modes and principal components (PC), which allow us to divide the variability in key parameters between spatial and temporal components. Each mode describes a certain fraction of a dispersion of the initial data. This fraction is inversely proportional to the order of a mode [*Hannachi et al.*, 2007]. The first 3-5 modes describe most of the dispersion of the analyzed salinity fields, which allows significant compressing of the information contained in the original data [*Hannachi et al.*, 2007; *Borzelli and Ligi*, 1998]. The EOF decomposition was carried out for the average salinity fields for the layer 5-50 m, as well as obtaining time series of PCs for the periods of 1950-1993 and 2007-2012.

We applied our statistical model to interpret the physical processes through PCs. We approximated the time series of principal components to identify predictors that determined variability of the salinity fields; also, it will help to obtain the equations for projection of future changes. The statistical model is presented by a system of linear regression equations constructed for the first five PCs, as the first five EOF yields are above 77 % of the total variance of the salinity data. The principal components were associated with these factors: the atmospheric circulation indexes (AMO, AO, NAO, PDO, PNA, DA; see Table 1), water exchange with the Pacific and Atlantic Oceans, river runoff, and the area of the ice-free surface in the Arctic seas in September. We should note here we did introduce one assumption that the time series of the Arctic and Atlantic Ocean water exchange can be presented through AMO indexes.

### 2.4. Phase portrait and cluster analysis

We used phase portraits to analyze the trajectories of the evolution of winter salinity. The points on the Figure 2 represent scores on the first two PCs for each of the 49 years. The phase portrait shows that the inter-annual oscillations in salinity have existed in a narrow range. The concentration of numerous points of the phase portrait is in the positive plane of the principal component values.

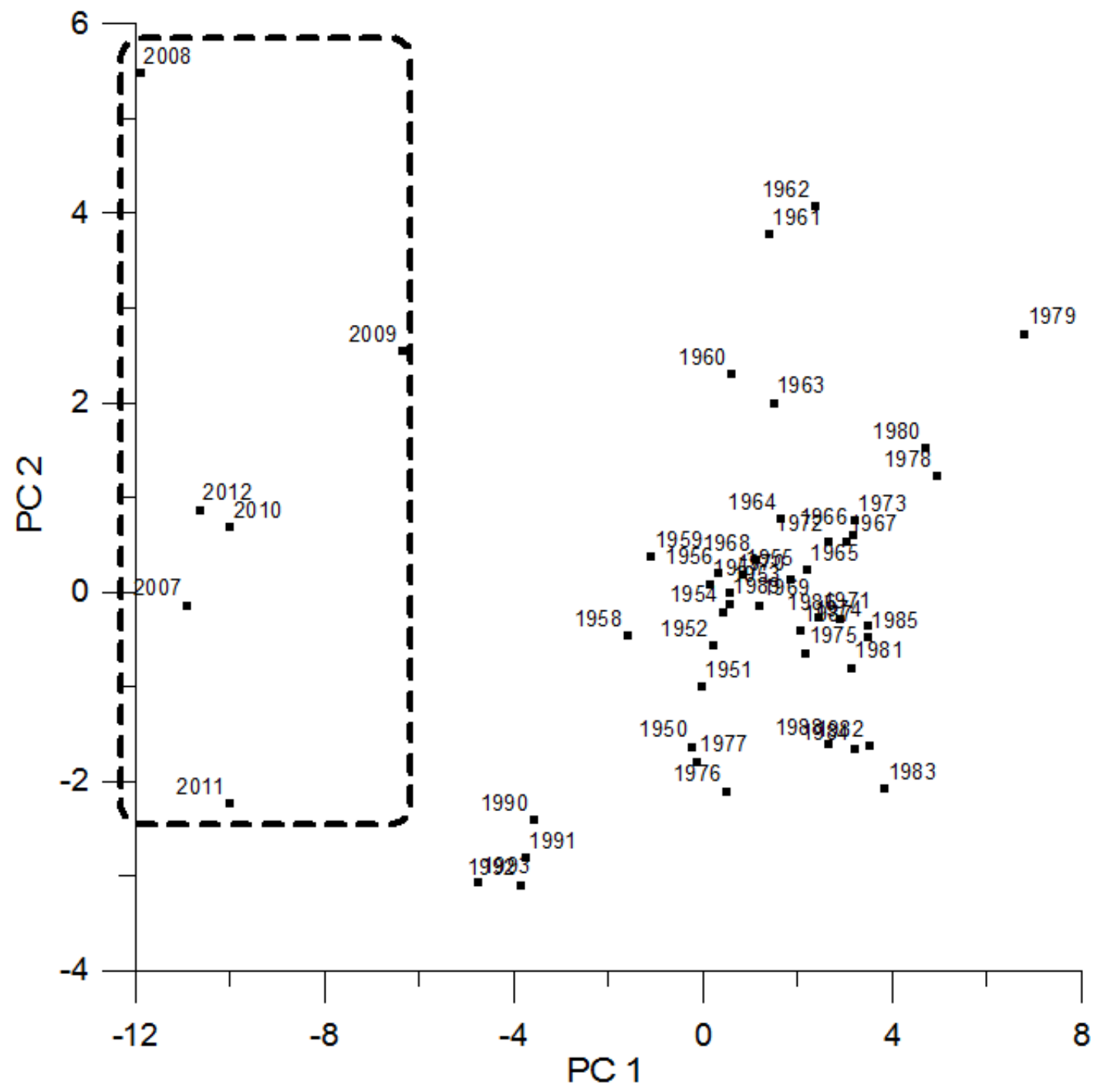

*Figure 2. Phase portrait for average salinity in the upper 5-50 m layer.*

We should note that 2007 was a “tipping” year, when the spatial structure of salinity in the Arctic Ocean sharply changed. The points of the phase portrait for the period 2007-2012 were allocated in the negative plane of the first principle component. In addition, these points are too far from the compact aggregation of the points on the positive plane for the period 1950-1989. Likely, the period 1990-1993 may be designated as a transition period. At this time the values of the first principle component were negative, but the spatial structure of the fields had just started a tendency to change. As we mentioned above, the points of the phase portrait characterizing the changes in the structure of the winter salinity fields for the period 1950-1989 can form the compact aggregation of the points. The goal of the analysis of this structure was to identify the patterns, but it was not that easy. The cluster analysis suggested by *Ward* [1963] should be really useful for investigation of the dynamics and anomalies in these patterns. According to this approach, we represented the salinity fields as a grid with nodes. Each of these nodes contained information about salinity in the region, and the measure of the distance between two nodes was introduced through the Euclidean metric:

$$D_{ij} = \left(\sum_{ij} (S_i - S_j)^2\right)^{\frac{1}{2}} \qquad (5)$$

where $S_i$ and $S_j$ are the values of the salinity in two nodes $i$ and $j$ over time.

Consequently, this analysis allowed us to obtain the hierarchical salinity fields with a feature of statistical identity (Figure 3). The figure shows that the salinity fields have structural differences and thus are grouped in clusters for consecutive years. Based on the tree ties, we have identified six of the largest groups in a temporal scale as well as the six basic patterns in salinity fields. The first cluster reproduces the field for the years: 1950-59, 1976-77 and 1989; the second cluster includes 1960-1965; the third cluster includes 1966-1975; the fourth cluster includes 1981-1988; the fifth cluster includes 1978-1980; and the sixth cluster includes 1990-93 and 2007-2012. All these clusters may be found on the phase portrait as well (Figure. 2).

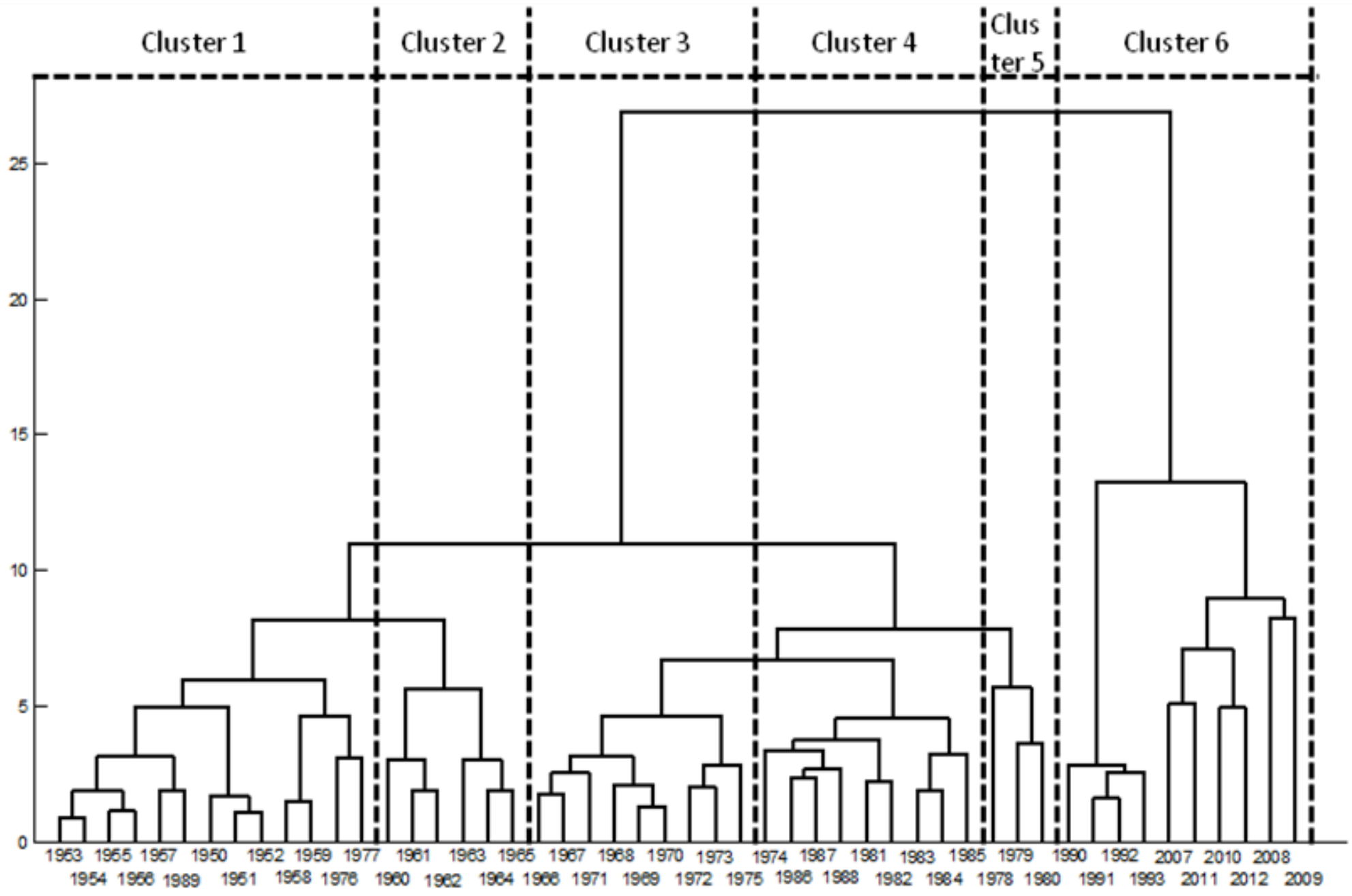

*Figure 3. Dendrogram of winter salinity fields for the layer 5-50 m in the Arctic Basin.*

Similar results were obtained using other methods of cluster analysis (e.g., complete linkage, and weighted pair-group average). This shows that the chosen division into clusters is stable and proper. In addition, similar dendrograms were found in the work of *Koltyshev et al.* [2008], where cluster analysis was completed for the data series of an average salinity at a depth of 5-25 m for the period of 1950-1989. It also confirms the robustness of our classification. Within the framework of our classification, the pattern may persist for two to nine years.

We calculated the average salinity fields for each period of each group. It allowed us to find the differences (from cluster to cluster) in the structure of salinity fields (Figure 4).

*Pattern 1:* Our analysis for these years shows that a desalination zone occupies the southern part of the Canada Basin (Figure 4a). The salt-frontal zone lies along the Lomonosov Ridge. This kind of distribution of salinity fields is formed under the dominance of a cyclonic mode of atmospheric circulation [*Proshutinsky and Johnson*, 1997].

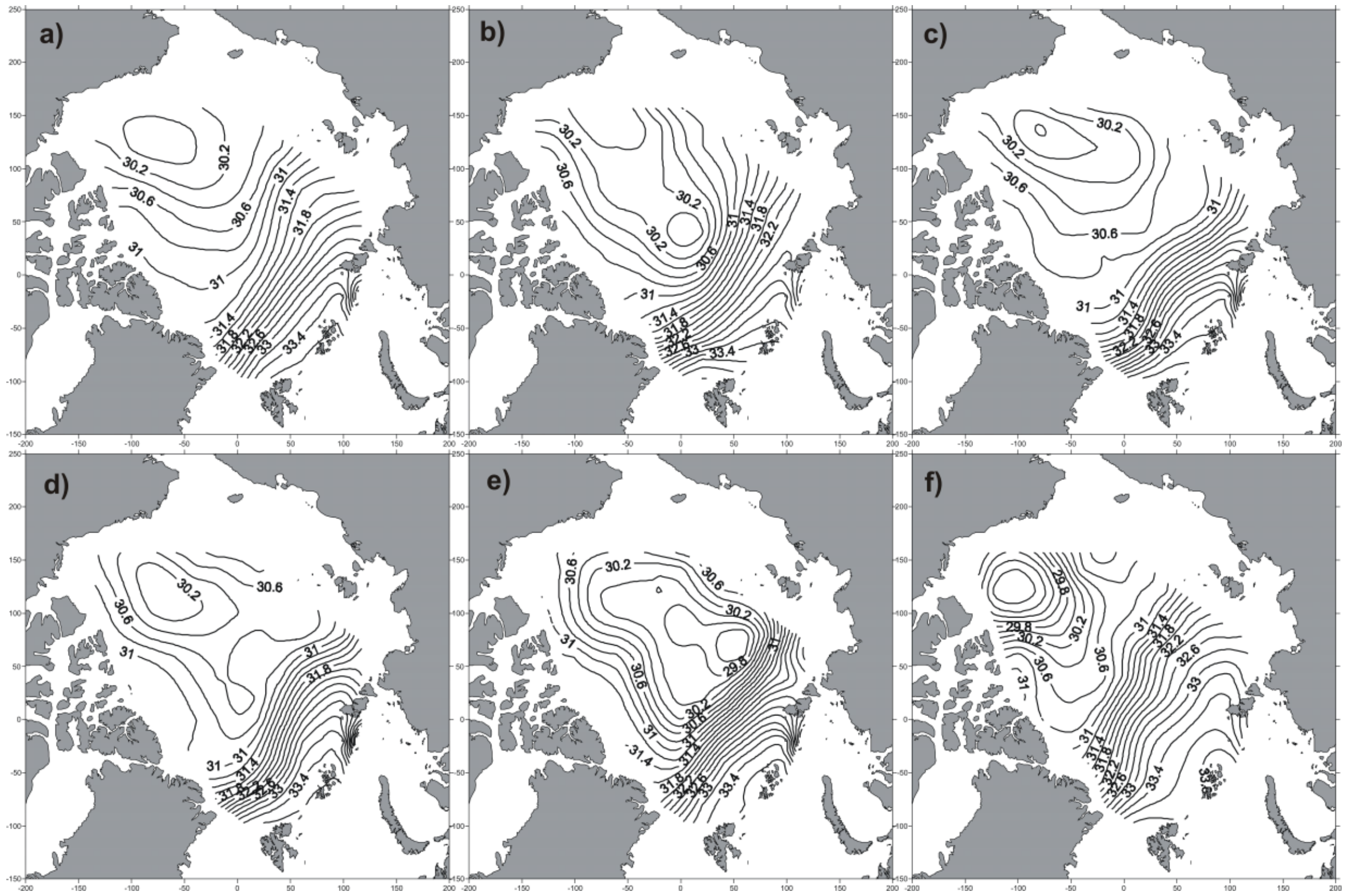

*Figure 4. Winter salinity fields for the layer 5-50 m averaged over periods to patterns: a – pattern 1; b – pattern 2; c – pattern 3; d – pattern 4; e – pattern 5; f – pattern 6.*

*Pattern 2:* Here the distribution of salinity fields look more like a freshening zone with multiple cores, which extend from the Beaufort Sea to the North Pole (Figure 4b). This structure of the spatial distribution of salinity is formed because of the anti-cyclonic mode of the atmospheric circulation at the different positions of the anti-cyclonic core.

*Pattern 3:* The main feature of the salinity distribution is an extensive area of freshening which occupies the entire Canada Basin. As a result, the salinity frontal zone is shifted to the region of the Gakkel Ridge (Figure 4c). This structure of the salinity spatial distribution is formed at the anti-cyclonic mode of the atmospheric circulation.

*Pattern 4:* We can see here that this cluster combines the salinity fields with a tendency to the formation of several zones in the prefrontal area of desalination, which is moving into the area of the Gakkel Ridge (Figure 4d).

*Pattern 5:* The core of freshening has a displacement to the region of the Makarov Basin to the Northeast from the slope of the Laptev Sea shelf. The freshening zone extends from west to east (Figure 4e).

*Pattern 6:* The zone of maximum freshening is located near the center of the Canada Basin. Also, this zone is connected to the freshening zone in the Beaufort Sea. Additionally, we can see the formation of a small core of freshening close to the region which is north of the East Siberian Sea. The salt-frontal zone occupies the extreme eastern position, lying on the Makarov

Basin (Figure 4f). This kind of salinity distribution is formed mainly under the influence of highly developed cyclonic atmospheric circulation.

In addition, we note that pattern 6 has a separate branch with the largest Euclidean distance on the dendrogram. Therefore, since 1990 the structure of the salinity fields has undergone significant changes, which were most pronounced in 2007-2012. These years can be isolated in a separate sub-branch.

Figure 5 illustrates the differences between patterns allocated previously for classification of surface field salinity in terms of PCs. Patterns 1 and 6 are characterized by negative values of the three principal components; the difference between the patterns are the amount of values for the first principal component ($PC_1$). Patterns 3, 4, and 5 are characterized by dominant positive values for $PC_1$ and different sign and magnitude of $PC_2$ and $PC_3$. Pattern 5 is the opposite of pattern 6 in terms of the PC values. As we see from Figure 4 (e, f), a shift in the signs of the principal components can be explained by moving the core of freshening from the Makarov Basin to the Beaufort Sea, and the degree of freshening appears to determine the absolute value of $PC_1$.

In general, if we compare the variability of salinity in the Eurasian and Canada Basins, we may conclude that the main difference in salinity fields for 2007-2012 (included in pattern 6) are in the value of the salinity of the Canada Basin. During this period its variance was less than 0.8 ‰ compared with average values. This means there has been a significant freshening of the surface layer, which had apparently not happened at least the past 50 years in available observations.

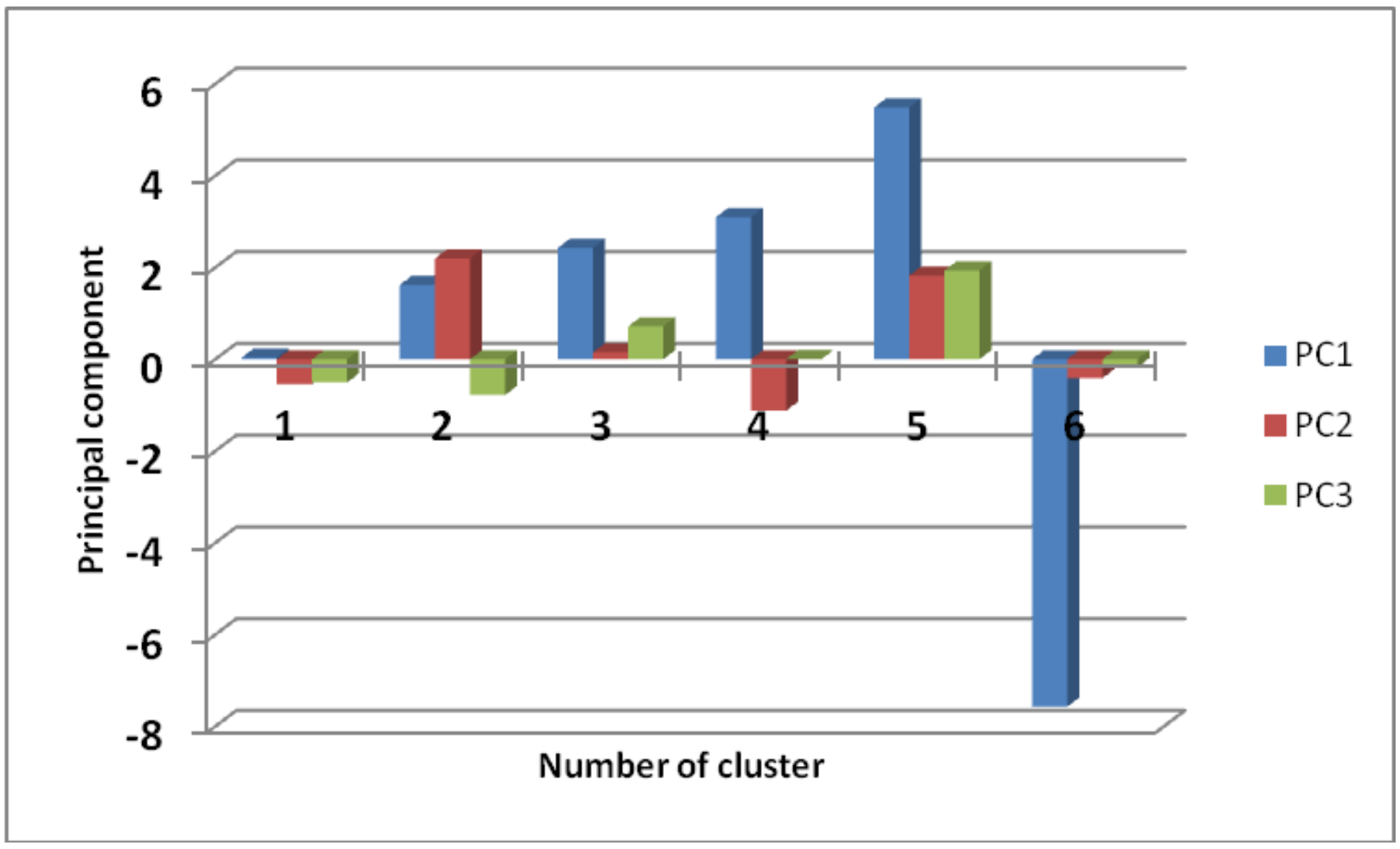

*Figure 5. The average values of PCs for six patterns: 1 - 1950-59, 1976-77 and 1989; 2 - 1960-1965.; 3 – 1966-1975; 4 – 1981-1988; 5 – 1978-1980.; 6 – 1990-93 and 2007-2012.*

**2.5. Decomposition of surface salinity fields by EOF**

Considering the results of cluster analysis, as a result of EOF decomposition of the average salinity fields for the 5-50 m layer, we obtained two sets of modes and principal components for the period of 1950-1993 (series 1), and for the same period adding the years 2007-2011 (series 2). In summary, the first three modes obtained by the decomposition of series 1 describe over 60% of the total dispersion of the initial fields. The first three modes of series 2 describe almost 67.5% of the total dispersion. These modes for both decompositions are significantly different (Figure 6).

We can see that the first mode has an additional core in the Canada Basin; we observed reorientation of the cores for the rest of the modes. The first mode of series 1 describes 38% of the total salinity variability, and the first mode of series 2 takes into account 51.5% of the initial data dispersion. The first mode is associated with the influence of large-scale atmospheric circulation in the Arctic [*Timokhov et al.*, 2012]. Therefore, we can conclude that the role of atmospheric circulation in the formation of the surface salinity fields in the Arctic Basin have grown significantly over the last decade. Thus, the modes obtained by decomposition in series 1 cannot take into account the essential features of the distribution of surface salinity fields associated with the freshening waters of the Canada Basin. Therefore, for further analysis we used the principal components and modes obtained upon decomposition in series 2.

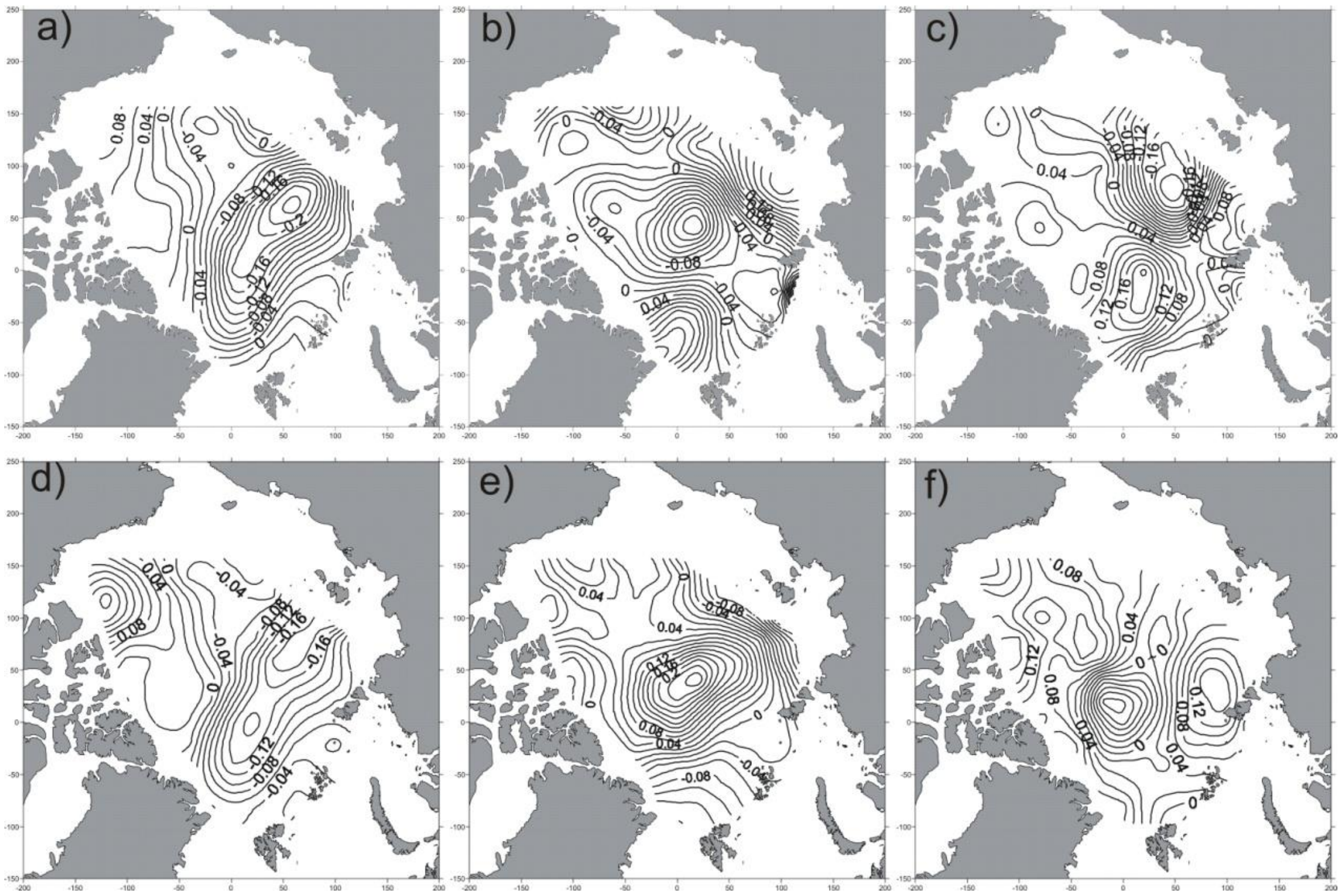

*Figure 6. The first three modes of the average salinity field decomposition for the layer 5-50 m: a, b, c - 1st, 2nd and 3rd modes, respectively, for the period 1950-1993.; d, e, f - 1st, 2nd and 3rd modes, respectively, for the period 1950-1993 and 2007-2011.*

### 2.6. The linear regression equation for the principal components

A set of external factors having the most correlation with the main components of salinity decomposition have been defined based on the results of correlation analysis. As a result of linear regression analysis we obtained empirical equations for the first five principal components (see Appendix).

The structure of these equations can be explained through the sets of factors that simulate the effects of both atmospheric and hydrological processes. Thus, the predictors used can be divided into two groups: the first group includes atmospheric circulation indices and air temperature anomalies, and reflect the influence of atmospheric processes; the second group – river runoff for Arctic seas, inflow through the Bering Strait, the relative areas of open water and ice in the Arctic in September – which are hydrological processes. Predictors were included in the equations with different time shifts. The value of the time shifts was 1–12 years.

The contribution of each group to the variability of $PC_1$ - $PC_5$ depends on the magnitude and sign of the predictors included in that particular group. In this case, hydrological processes have a dominant influence on $PC_1$ (in a ratio of 60/40%) and, vice versa, atmospheric processes are the major factors influencing $PC_2$ and $PC_3$ in the same proportion. Atmospheric and hydrological processes make approximately the same contribution (in a ratio of 47/53%) to the formation of the inter-annual variability of $PC_4$. For $PC_5$ the impact of these processes is assessed in relation 43.5/56.5%.

## 3. Discussion and Summary

In the late 1980s, the atmospheric circulation regime began to change [*Steele and Boyd*, 1998; *Kurazhov et al.*, 2007; *Proshutinsky, et al.* 2009; *Morison et al.,* 2012]. Degradation of the Arctic anticyclone is an excellent example of this change. Some changes in the structure of the surface pressure field were observed. This happened because of a frequent recurrence of large values of the DA indices.

According to *Wang et al.* [2009] this could be a main reason for the local minima of sea ice extent in the summers of 1995, 1999, 2002, 2005 and 2007. In addition, in the late 1980s the inflow of warm and high salinity Atlantic water into the Arctic Basin increased [*Frolov et al.*, 2009]. At the beginning of this century, the heat flow of Pacific waters through the Bering Strait to the Chukchi Sea increased [*Woodgate et al.*, 2010].

We calculate correlations of the principal components with different climate processes such as the atmospheric processes, river runoff, and the volume of water coming in through the straits of the Arctic Basin (Table 1). Statistically significant coefficients were obtained for factors reflecting influences on the processes listed above. Thus, we can assume that *Pattern 6* of the dendrogram is the consequence of these processes.

The time series of some of these processes have been normalized over the interval 0 to 1. We chose the clusters (1950-59, 1976-77, 1989 (*Pattern 1*) and 1990-1993, 2007-2012 (*Pattern 6*)) with a similar structure of their surface salinity fields (Figure 4a and 4f), but with different values of salinity in the water cycle of the Beaufort Sea.

The histogram (Figure 7) shows that the relative values of almost all factors for the years 1990-1993 and 2007-2012 were significantly higher than in the year 1950. Temperature anomalies, the area of ice-free regions of the shelf seas, winter and summer AO indexes and DA indexes have reached their highest values.

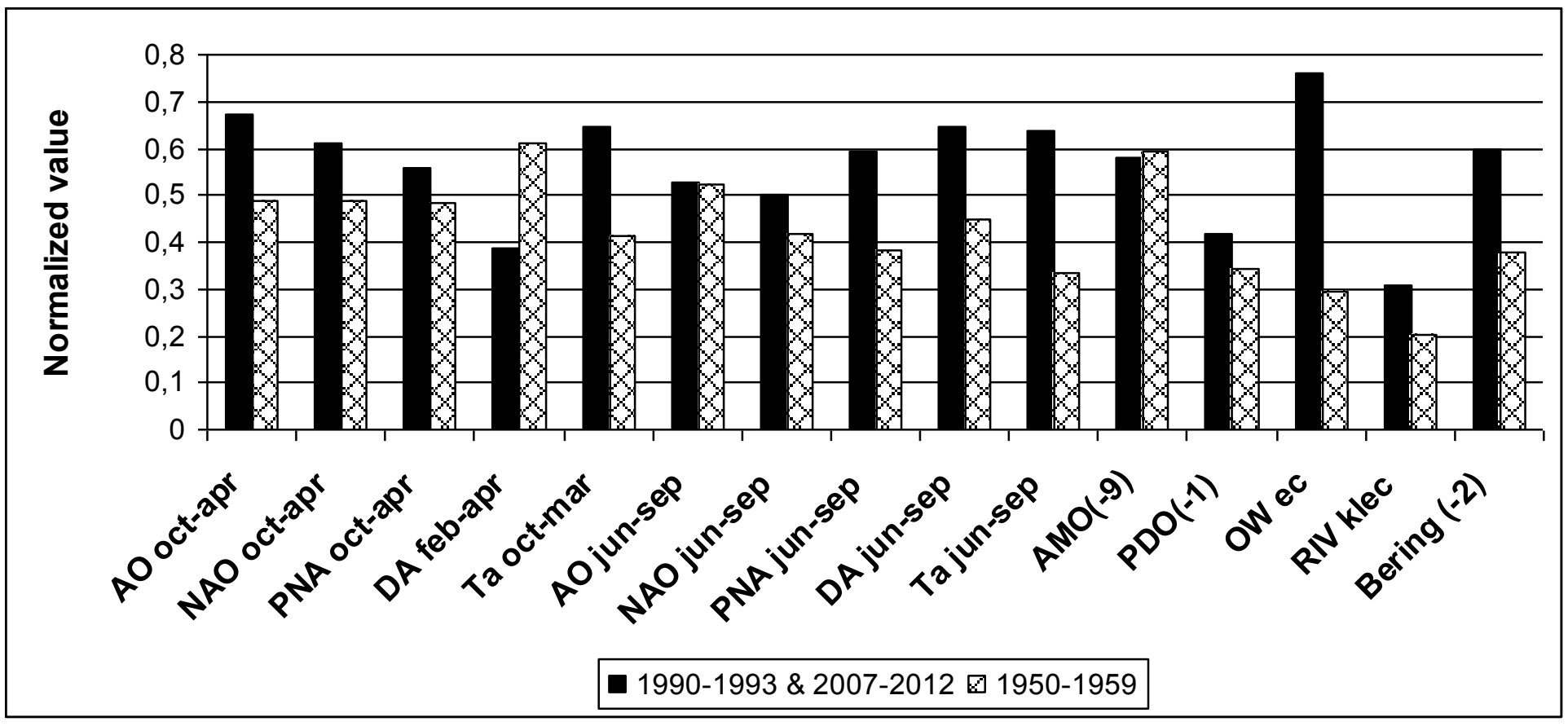

*Figure 7. Average values of the normalized environmental factors (see Table 1). Indices of atmospheric circulation and temperature anomalies which, averaged over the winter and summer months, have been used in the calculations.*

We present here a statistical model of inter-annual variability of the Arctic Ocean surface layer salinity. This research builds on already established approaches used by *Pokroivsky and Timokhov* [2002] (specifically, their reconstruction of salinity fields applying modified EOF methods) and *Timokhov et al.* [2012].

However, our contribution to their work is the formulation of a statistical approach, which can be used as a universal tool for analysis of inter-annual variability of Arctic Ocean surface layer salinity. Moreover, we suggest some additional things to improve the ideas presented in previous research. For example, as opposed to the research of *Timokhov et al.* [2012], we do not take into account the previous values of the principal components (history) which simplifies the calculations and allows us to consider earlier data. In addition, we also make calculations using current observational data, which is quite important for understanding the physical processes during the dramatic current changes in Arctic sea ice.

The derived equations in the Appendix describe the first five principal components for the period 1950-2014; PCs for 1950-2011 obtained from these equations have good agreement with the values of PCs directly derived from the decomposition of salinity fields on EOF (Figure 8).

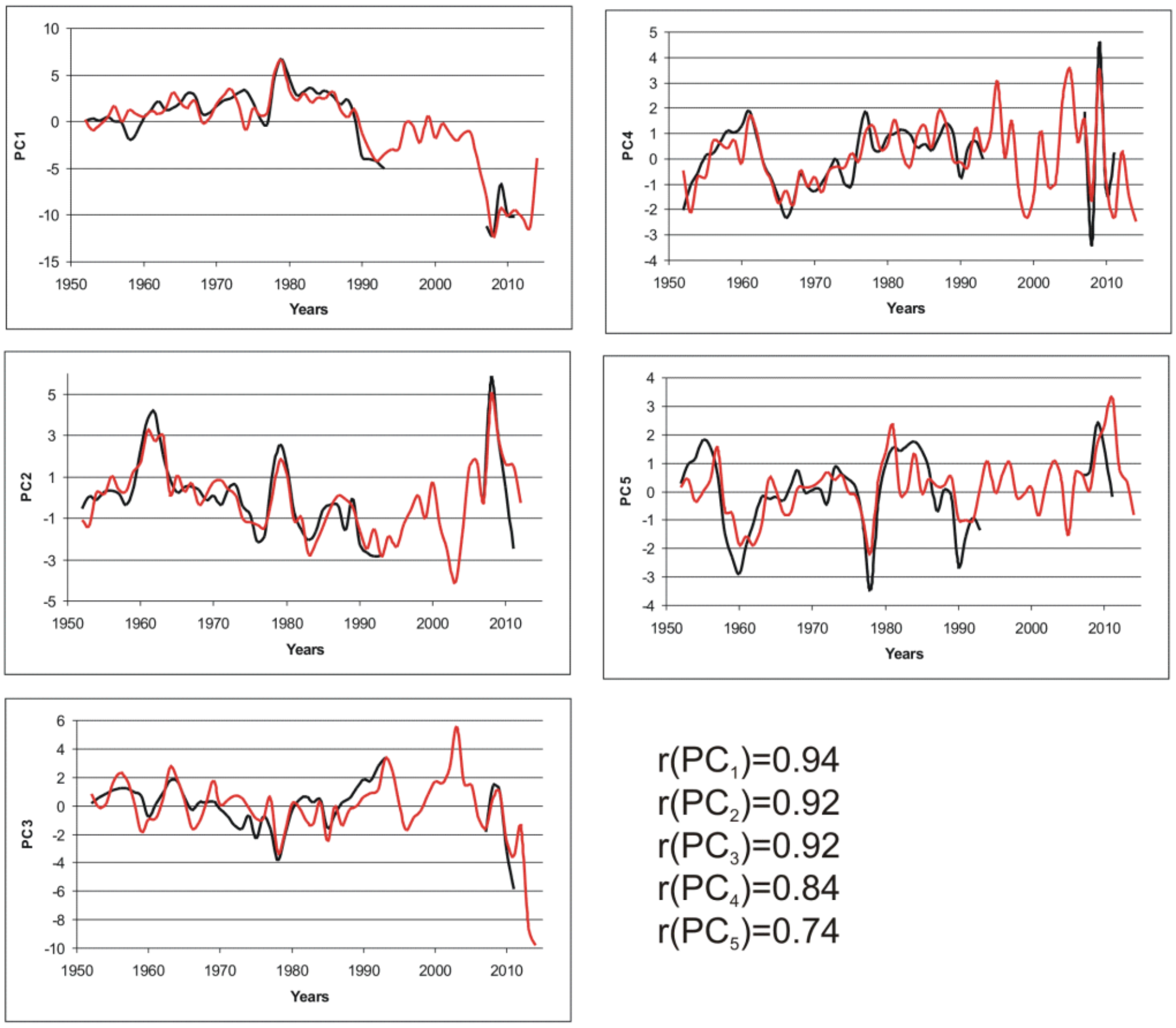

*Figure 8. The actual (black line) principal components and calculated principal components (red line) with the help of the equations of linear regression. Also, we show the correlation coefficients r(PC$_i$) between the calculated time series of PCs and the real PCs, obtained by decomposition of the salinity fields on EOF.*

Theoretically, the salinity fields for 1994-2006 can be reconstructed with the help of this model. We noted above that this period had gaps in observational data. But in practice it may lead to inaccuracies due to the higher-frequency variability of calculated PCs. This model cannot reproduce exact principal components for short-term time series, although the trend in variability of all five PCs is reproduced correctly. Therefore, the model can be used for tracking long-term processes of the structure transformation of salinity fields. Using this useful tool we can make projections for anomalies, their frequency, and ultimately approach an understanding of these sophisticated physical processes.

Validation of the model was carried out by calculating an error of reconstruction for the surface salinity fields. The difference between the actual and reconstructed salinity fields was determined as a percentage by the following formula:

$$I_{rm} = (\sigma(S_f - S_c)/\sigma(S_f)) \cdot 100\%, \qquad (6)$$

where $\sigma$ is the standard deviation; $S_f$ is the actual salinity, and $S_c$ is the calculated salinity. The error in the reconstruction of salinity fields is 25.2 % (Figure 9). There are several possible reasons for this.

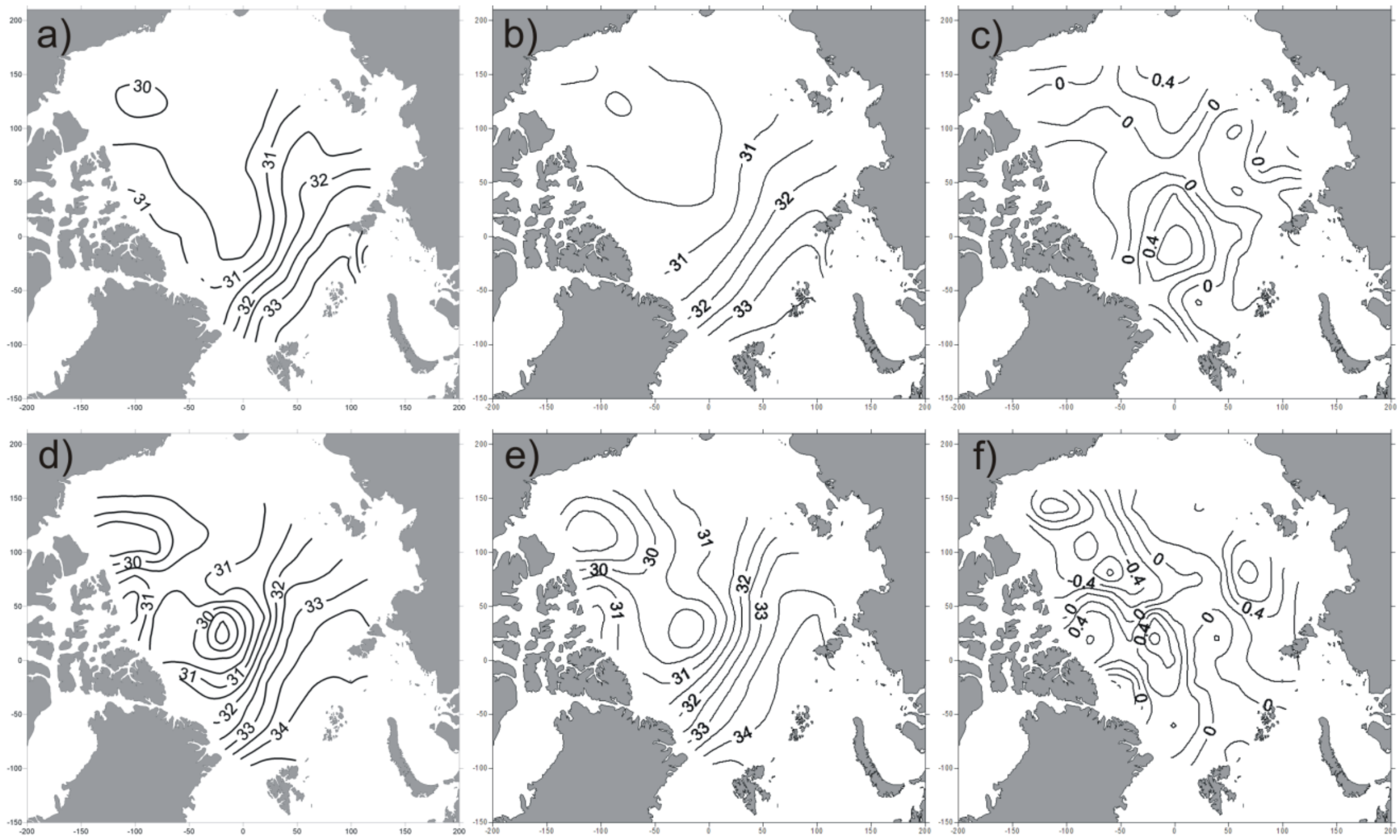

*Figure 9. The real average salinity field for the layer of 5-50 m (a, d), the reconstructed average salinity fields for the layer of 5-50 m (b, e) and the difference between these fields (c, f) for 1955 (upper line) and 2009 (bottom line).*

The first five EOF modes describe more than 77 % of the variability of the initial fields. It is possible that the characteristics of salinity fields may reproduce the higher order modes [*Borzelli and Ligi*, 1998]. If the order of a mode increases, then the dispersion decreases. So, it can enhance uncertainty in interpreting the physical processes associated with PCs. Thus, the error of reconstruction in salinity fields initially incorporated into the model is about 23%.

During the last decade, there have been significant changes in the thermohaline state of the surface layer. Some authors say there is a complete restructuring of the thermohaline structure of the surface layer and its transition to a qualitatively new state [*Timokhov et al.*, 2011]. It is possible that we are using modeling factors that cannot fully describe currently observed changes. For example, we could not find sufficiently long data series of river flow into the Beaufort Sea. But, in this area significant freshening of the surface layer is observed and the largest reconstruction errors occurred (Figure 9).

Also, we applied this model to the reconstruction of salinity fields for 2013-2014. It should be noted that the time series of some predictors were insufficient in length for getting values of PCs. Therefore extrapolation was made.

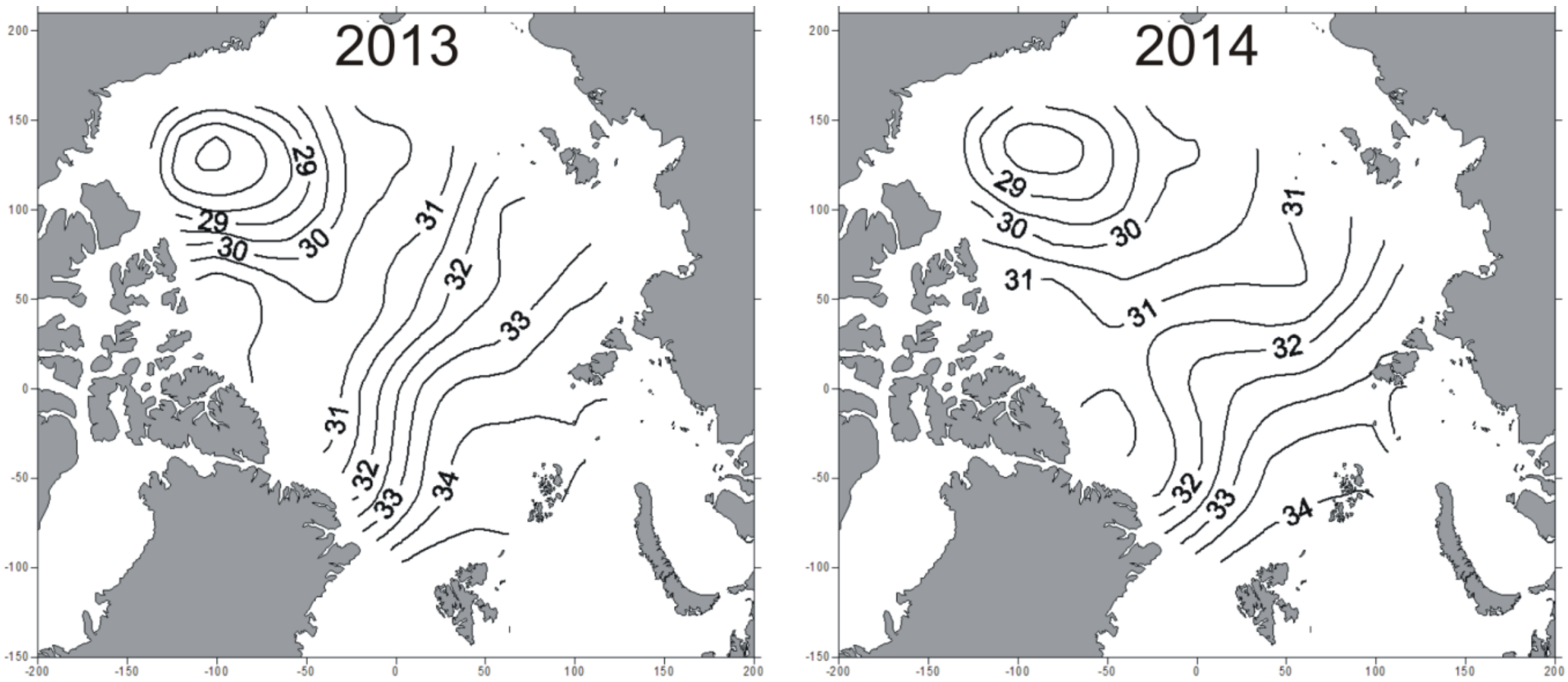

*Figure 10. There is reconstructed salinity field for the layer of 5-50 m in 2013 and 2014.*

As a result, we obtained the salinity field which corresponds to the observed trends in recent years. This has preserved the significant freshening in the Canada Basin as well as large spatial gradients between the Eurasian and Canada Basins. According to our projections for 2013-2014 (Figure 10), freshened water from the Beaufort Gyre will move westward along the Siberian Continental Slope. In 2014, the spatial structure of the salinity field is similar to the structure that is typical for fields belonging to *Pattern 4* (1981-1988), but they differ by the surface salinity values in the Beaufort Sea.

Thus, here we have identified various patterns in the Arctic Ocean surface salinity fields with the help of a reliable statistical model. In addition, we have found anomalies in the salinity fields which have occurred in the past. Our findings again raise the question about nonlinearities in global ocean circulation, and particularly in the Arctic Ocean which is strongly connected with the Arctic climate system. In the future, information obtained about these anomalies may be helpful in determining whether the Arctic Ocean salinity, and related oceanographic phenomena, have reached a tipping point or critical threshold.

**Acknowledgments**

The authors thank the big data sources, which are mentioned below.

Detailed algorithm of salinity data gridding procedure is available at the AARI web-site: http://www.aari.ru/resources/a0013_17/kara/Atlas_Kara_Sea_Winter/text/tehnik_report.htm#p2. Indexes of atmospheric circulation are available at NOAA's Data Base and JISAO Climate Data Archive. The area of the ice-free surface in the Arctic seas was calculated from data sited at http://www.aari.ru/projects/ECIMO/index.php?im=100. Ice extent was recalculated from ice concentration data that are available at NOAA's Gridded Climate Datasets: Ocean. River runoff data were obtained from the Joint US-Russian Atlas of the Arctic Ocean and ArcticRIMS Data Server. Also, we are grateful for financial support from the Otto Schmidt Laboratory for Polar and Marine Research (Grant No. OSL-13-05), and from the Russian Foundation for Basic Research (Grant No. 14-01-31053). Finally, we gratefully acknowledge support from the Division of Mathematical Sciences and the Division of Polar Programs at the U.S. National Science Foundation (NSF) through Grants ARC-0934721, DMS-0940249, and DMS-1413454. We are also grateful for support from the Office of Naval Research (ONR) through Grant N00014-13-10291. We would like to thank the NSF Math Climate Research Network (MCRN) as well for their support of this work. In preparing this text, we have benefited from discussions with Jessica R. Houf.

**Appendix: The empirical equations for the first five principal components**

The equations have proceeded from a simple formula of multiple linear regressions:

$$y_i = \sum a_{ij} x_{ij} + b_i$$

(A1)

where $y_i$ – the principal components $PC_i$; $x_{ij}$– the independent variables of $y_i$ (the different environmental factors), $a_{ij}$ – regression coefficients, $b_i$ – intercept.

Values of correlation coefficients (R), coefficients of determination ($R^2$) and F-criteria [*Hill and Lewicki*, 2007] are shown in Table A1. The values of all F-criteria exceeded the allowable limit, indicating that variables used are statistically significant. The correlation coefficients for all PCs were quite big, significant and varies from0.74 ($R_5$) to 0.93 ($R_1$).

Table A1. The empirical equations are determining PCs.

| **$PC_i$** | **Statistical Equations** | **Multiple R** | **Multiple $R^2$** | **Adjusted $R^2$** | **F-criteria** |
|---|---|---|---|---|---|
| $PC_1$ | $PC_1 = -0.96 \times \mathrm{AO}_{I-IV}(-2)^{*} - 1.11 \times \mathrm{AO}_{I-IV}(-1) - 1.62 \times \mathrm{NAO}_{XII-IV}(-1) - 3.17 \times \mathrm{AMO}(-8) - 7.38 \times \mathrm{BS}(-3) - 0.01 \times \mathrm{RIV}_{\mathrm{EC}}(-3) + 0.003 \times \mathrm{RIV}_{\mathrm{KL}}(-5) - 0.003 \times \mathrm{OW}_{\mathrm{KLEC}}(-1) + 9.53$ | 0.93 | 0.88 | 0.85 | 33.00 (8;37)** |
| $PC_2$ | $PC_2 = -0.57 \times \mathrm{AO}_{I-IV}(-1) - 1.49 \times \mathrm{AO}_{VII-IX}(-1) + 6.76 \times \mathrm{AMO}(-10) + 0.88 \times \mathrm{PDO}(-3) - 0.71 \times \mathrm{PDO}(-10) - 3.09 \times \mathrm{BS}(-4) - 0.006 \times \mathrm{RIV}_{\mathrm{LE}}(-3) - 0.005 \times \mathrm{RIV}_{\mathrm{EC}}(-5) + 0.003 \times \mathrm{OW}_{\mathrm{KLEC}}(-1) + 6$ | 0.91 | 0.84 | 0.79 | 20.64 (9;36) |
| $PC_3$ | $PC_3 = -0.68 \times \mathrm{NAO}_{XII-IV}(-3) + 7.65 \times \mathrm{AMO}(-5) - 3.53 \times \mathrm{AMO}(-8) - 2.42 \times \mathrm{AMO}(-9) + 3.42 \times \mathrm{AMO}(-11) + 1.40 \times \mathrm{PDO}(-10) + 6.44 \times T_{air(II-IV)}(-1) - 5.80 \times \mathrm{BS}(-3) + 0.002 \times \mathrm{RIV}_{\mathrm{KLEC}}(-3) - 0.002 \times \mathrm{RIV}_{\mathrm{KLE}}(-5) - 0.006 \times \mathrm{RIV}_{\mathrm{LE}}(-6) - 0.001 \times \mathrm{OW}_{\mathrm{KLEC}}(-1) + 13$ | 0.92 | 0.85 | 0.80 | 16.10 (12;33) |
| $PC_4$ | $PC_4 = 0.78 \times \mathrm{NAO}_{XII-IV}(-1) + 0.59 \times \mathrm{PNA}_{X-IV}(-1) - 0.60 \times \mathrm{PNA}_{VII-IX}(-1) + 2.79 \times \mathrm{AMO}(-6) - 2.18 \times \mathrm{AMO}(-12) - 0.66 \times \mathrm{PDO}(-6) - 8.27 \times \mathrm{BS}(-4) - 0.006 \times \mathrm{RIV}_{\mathrm{LEC}}(-6) + 0.001 \times \mathrm{OW}_{\mathrm{KLEC}}(-1) + 0.002 \times \mathrm{OW}_{\mathrm{EC}}(-1) + 8.83$ | 0.83 | 0.70 | 0.62 | 8.27 (10;35) |
| $PC_5$ | $PC_5 = -0.68 \times \mathrm{NAO}_{I-IV}(-1) - 2.38 \times \mathrm{AMO}(-7) - 3.52 \times \mathrm{AMO}(-12) + 4.72 \times T_{air(II-IV)}(-2) + 0.001 \times \mathrm{IceExt}(-1) - 0.002 \times \mathrm{RIV}_{\mathrm{KL}}(-5) + 0.007 \times \mathrm{RIV}_{\mathrm{LEC}}(-6) + 0.001 \times \mathrm{OW}_{\mathrm{KLEC}}(-2) - 11.74$ | 0.74 | 0.55 | 0.46 | 5.70 (8;37) |

* – time shift for every predictor is indicated in brackets (minus means that predictor outpace dependent variable).

** – numbers of degrees of freedom are shown in brackets.